\documentclass[%
reprint,
amsmath,amssymb,
aps,
pra,
]{revtex4-2}

\usepackage{graphicx}
\usepackage{dcolumn}
\usepackage{bm}

\begin{document}


\title{Special Theory of Relativity for a Graded Index Fibre}

\author{Shinichi Saito}
 \email{shinichi.saito.qt@hitachi.com}
\affiliation{Center for Exploratory Research Laboratory, Research \& Development Group, Hitachi, Ltd. Tokyo 185-8601, Japan.}

\date{\today}

\begin{abstract}
The speed of light ($c$) in a vacuum is independent on a choice of frames to describe the propagation, according to the theory of relativity.
We consider how light is characterised in a material, where the speed of light is different from that in a vacuum due to the finite dielectric constant.
The phase velocity in a material is smaller than $c$, such that the speed of a moving frame can be larger than the phase velocity, such that the frame can move faster than the speed of light in a material.
Consequently, an unusual Doppler effect is expected, and the wavelength in the moving frame changes from the red-shift to the blue-shift upon increasing the speed of the frame.
The corresponding energy of the light also changes sign from positive to negative, while momentum is always positive, leading to the changes of sings for the phase velocity and the helicity.
In a graded index fibre, where the exact solution is available, even more complicated phenomena are expected, due to the finite effective mass of photons.
Upon the increase of the energy gap, generated by optical confinements and optical orbital angular momentum, the effective mass of photons increases.
If the gap is large enough, momentum starts to change the sign upon increasing the frame velocity, while the energy of photons is always positive. 
In this case, the phase velocity diverges if momentum is in agreement with the fame velocity.
Contrary to the unusual behaviours of the phase velocity, the group velocity is always below $c$.
This thought-experiment might be useful to consider the insight for the polarisation sate of light.
\end{abstract}

\maketitle


\section{Introduction}
Einstein made a legacy for the establishment of the theory of relativity \cite{Einstein05,Einstein05b}, which continues to attract a wide range of researchers and engineers for more than a century.
It was told that Einstein considered how light is seen from an observer travelling as fast as the speed of the light \cite{Lehner14}, leading to the discovery of the law, that the speed of light ($c$) is independent on a choice of a frame to describe propagation of light in a vacuum, incurred by Maxwell equations \cite{Jackson99,Yariv97}.
The universal relationship of space and time through the Lorentz transformation led various non-trivial results, such as Doppler effects of light, time dilation, length contraction, and the energy-momentum relationship of $E^2=(cp)^2+(m_0 c^2)^2$, where $E$ is energy, $p$ is momentum, $m_0$ is the rest mass of an object \cite{Einstein05,Einstein05b,Jackson99,Garetz81,Nienhuis96}. 
In order to satisfy the causality, it is strictly forbidden to allow motion faster then $c$  \cite{Einstein05,Einstein05b,Jackson99}.

Nevertheless, inside a material, a speed of an moving frame can exceed the speed of light due to the larger refractive index ($n$) than 1 of a vacuum, as experimentally proved by Cherenkov radiation \cite{Cherenkov34,Frank37,Cherenkov37,Cherenkov37b,Cherenkov37c,Cherenkov86,Baryshevsky17}.
A charged particle with the speed exceeding the phase velocity of light in a material produces a coherent shock-wave, similar to the sonic waves made by a supersonic aircraft.
Cherenkov radiation is usually observed in water as a bluish conical ray, emitted from a charged elementary particle, which enabled physicists to observe the neutorino oscillations \cite{Fukuda98}.
Here, we would like to revisit the original proposition of Einstein: {\it how the light is seen in a material from an observer, travelling at the speed exceeding the phase velocity of light?}

Our motivation is to understand the internal quantum structure of a photon, especially for getting insights for clarifying the correlation between spin and polarisation \cite{Saito20a,Saito20b,Saito20c,Saito20d}.
One might think that it is firmly well-established that spin of photons describes the polarisation of light \cite{Baym69,Sakurai14,Jackson99,Yariv97,Goldstein11,Gil16,Pedrotti07,Hecht17} such that the correlation is obvious.
However, this is highly non-trivial, as implied by Einstein, confessing that he could not understand what is a light quanta at all after 50 years of continuous considerations \cite{Lehner14}, regardless of the fact that he established a theory for a photoelectric effect as an evidence of the particle nature of a photon.

We have a hypothesis that polarisation of light is a macroscopic manifestation for spin of a photon as a quantum-mechanical feature.
We have shown that a wavefunction of a photon is described by a Helmholtz equation \cite{Saito20a,Jackson99,Yariv97}, which does not necessarily give a plane-wave form and the solution depends on a profile of the refractive index and a symmetry of the system.
A ray of photons, emitted from a laser source, is described by a many-body coherent state \cite{Grynberg10,Fox06,Parker05,Nagaosa99,Wen04,Altland10,Saito20a,Saito20b,Saito20c,Saito20d} with fixed phases, which describe an $SU(2)$ state for the spin state of macroscopically condensed photons.
By calculating the quantum-mechanical average of spin operators, we have shown that the expectation values of spin of photons are actually Stokes parameters in Poincar\'e sphere \cite{Stokes51,Poincare92,Saito20a}.
The magnitude of spin becomes $S_0=\hbar {\mathcal N}$, where $S_0$ it the Stokes parameter for the magnitude of polarisation, $\hbar$ is the Dirac constant, and ${\mathcal N}$ is the number of photons in a system, for a coherent ray of photons, which implies that the effective Planck constant becomes a macroscopic value, leading to a macroscopic realisation of a quantum state as polarisation \cite{Saito20a}.
The photonic orbital angular momentum is also a well-defined quantum-mechanical observable \cite{Allen92,Saito20b,Saito20c}, and we have shown that we can split spin and orbital angular momentum from the total optical angular momentum  \cite{Allen92,Enk94,Leader14,Barnett16,Yariv97,Jackson99,Grynberg10,Bliokh15,Enk94,Leader14,Barnett16,Chen08,Ji10,Leader14,Saito20c} 
in a GRaded-INdex (GRIN) fibre \cite{Kawakami68}, where the exact solution is available based on a Laguerre-Gauss mode \cite{Saito20c}.
Spin of a photon is derived by two-dimensional ($2D$) space-time Dirac equation from the principle of a rotational symmetry for a quantum-mechanical state of a photon \cite{Saito20d}.
Based on these considerations \cite{Saito20a,Saito20b,Saito20c,Saito20d}, we believe that the coherent spin state of photons from a laser source is characterised by a broken symmetry state due to the Bose-Einstein condensation of photons, enabled by pumping above the lasing threshold \cite{Saito20d}.
Consequently, the macroscopically coherent ray of photons from a laser source is described by a single $SU(2)$ wavefunction such as a Jones vector \cite{Yariv97}, a chiral Bloch state \cite{Baym69,Sakurai14}, and a diagonal state \cite{Saito20a}.
Thus, a simple quantum-mechanical calculation of a spin state is applicable to a coherent photonic ray using a single particle wavefunction, and its manipulation is also straightforward by employing a phase-shifter and a rotator to change the phases of the wavefunction \cite{Saito20a}.

The aim of this paper is to understand the photonic state in a material seen from an observer travelling at the speed comparable or even larger than the speed of light.
In a photonic crystal \cite{Joannopoulos08} or an optical fibre \cite{Yariv97}, the dispersion relationship of a photon is precisely engineered to adjust the speed of light and other photonic properties.
We consider a uniform material and a GRIN fibre as examples, because we can treat the dispersion, exactly in an analytic way.

\section{Principle}
\subsection{Lorentz transformation}
We are not challenging against the established Lorentz transformation at all \cite{Einstein05,Einstein05b,Jackson99,Weinberg05}. 
The energy scale, we are considering is of the order of 1eV such as for fibre optics, such that the space-time relationship of the vacuum must be robust in the presence of a material.
We assume the rest frame of $(t,x,y,z)$, where fibre optic materials are located, and consider how the light will be observed by the moving from of $(t^{\prime},x^{\prime},y^{\prime},z^{\prime})$ at the speed of $v_z$ along the positive $+z$ direction.
The Lorentz transformation, $L$, whose determinant is unity, $\det(L) =1$, is defined by
\begin{eqnarray}
\left (
  \begin{array}{cc}
   c t^{'}
\\
   z^{'}
  \end{array}
\right)
&=&
L
\left (
  \begin{array}{cc}
   c t
\\
   z
  \end{array}
\right)
\\
&=&
\left (
  \begin{array}{cc}
   a & -b \\
   -b & a
  \end{array}
\right)
\left (
  \begin{array}{cc}
   c t
\\
   z
  \end{array}
\right)
\\
&=&
\left (
  \begin{array}{cc}
   a c t - b z
\\
   - b c t + a z
  \end{array}
\right),
\end{eqnarray}
where $a^2-b^2=1$, and parameters $a$ and $b$ are determined by the principle of relativity, guaranteeing that the speed of light in a vacuum is independent on the measurement frame.
In order to impose the principle, we define the $2D$ d'Alembertian operators as 
\begin{eqnarray}
\Box_2
=
\partial_z^2-\frac{1}{c^2} \partial_t^2
=
\Box_2^{'}
=  
\partial_{z^{'}}^2-\frac{1}{c^2} \partial_{t{^{'}}}^2.
\end{eqnarray}
By inserting the identity

\begin{eqnarray}
\left (
  \begin{array}{c}
   \frac{\partial}{\partial t}
\\
\\
  \frac{\partial}{\partial z}
  \end{array}
\right)
=
\left (
  \begin{array}{c}
   \frac{\partial t^{'}}{\partial t}
   \frac{\partial }{\partial t^{'}}
+
   \frac{\partial z^{'}}{\partial t}
   \frac{\partial }{\partial z^{'}}
\\
\\
   \frac{\partial t^{'}}{\partial z}
   \frac{\partial }{\partial t^{'}}
+
   \frac{\partial z^{'}}{\partial z}
   \frac{\partial }{\partial z^{'}}
  \end{array}
\right)
=
\left (
  \begin{array}{c}
   a
   \frac{\partial }{\partial t^{'}}
-
   b c
   \frac{\partial }{\partial z^{'}}
\\
\\
   -\frac{b }{c}
   \frac{\partial }{\partial t^{'}}
+
   a
   \frac{\partial }{\partial z^{'}}
  \end{array}
\right),
\nonumber \\
\end{eqnarray}
into d'Alembertian, we obtain
\begin{eqnarray}
\frac{d z^{'}}{d t^{'}}
=
\frac{-b c + a \frac{dz}{dt}}{a-\frac{b}{c}\frac{dz}{dt}},
\end{eqnarray}
where $dz^{\prime}/dt^{\prime}=-v_z$ at $dz/dt=0$.
Then, we obtain $a$ and $b$, and $L$ becomes a standard form \cite{Einstein05,Einstein05b,Jackson99,Weinberg05} of
\begin{eqnarray}
L&=&
\frac{1}{\sqrt{1-\beta^2}}
\left (
  \begin{array}{cc}
1 & -\beta\\
-\beta & 1
  \end{array}
\right)
\\
&=&
\gamma
\left (
  \begin{array}{cc}
1 & -\beta\\
-\beta & 1
  \end{array}
\right)
,
\end{eqnarray}
where $\gamma=1/\sqrt{1-\beta^2}$ and the normalised velocity of the moving frame is given by $\beta=v_z/c$.
By using the Lorentz transformation, the d'Alembertian is always invariant in a vacuum, such that the velocity of light is constant and independent on the choice of the frame.
We can also consider the inverse Lorentz transformation as 
\begin{eqnarray}
\left (
  \begin{array}{cc}
   c t
\\
   z
  \end{array}
\right)
&=&
L^{-1}
\left (
  \begin{array}{cc}
   c t^{'}
\\
   z^{'}
  \end{array}
\right)
\\
&=&
\gamma
\left (
  \begin{array}{cc}
1 & \beta\\
\beta & 1
  \end{array}
\right)
\left (
  \begin{array}{cc}
   c t^{'}
\\
   z^{'}
  \end{array}
\right),
\end{eqnarray}
which is equivalent to exchange
\begin{eqnarray}
\beta &\leftrightarrow& -\beta \\
t &\leftrightarrow& t^{\prime} \\
z &\leftrightarrow& z^{\prime} 
\end{eqnarray}
in the original Lorentz transformation, reflecting the principle of relativity.
The Lorentz transformation is valid in our argument.

\subsection{Sch\"odinger equation for a photon}
In a material, the dispersion relationship for a photon is highly non-trivial.
In order to provide a specific example, we chose a GRIN fibre, where the exact solution is available \cite{Kawakami68,Yariv97,Saito20a,Saito20b,Saito20c,Saito20d}.
The refractive index of a GRIN fibre ($n$) is given by $n^2=n_0^2\left(1-(gr)^2\right)$, where $n_0$ is the refractive index of the core, $g$ is the graded index parameter, and $r$ is the radius in a cylindrical coordinate of $(r,\phi,z)$.
The radial and angular dependences of the wavefunction for a photon in a GRIN fibre can be decoupled by using the Laguerre-Gauss mode  \cite{Allen92,Yariv97,Saito20a,Saito20b,Saito20c,Saito20d}, which is equivalent to integrating over these degrees of freedom in a Feynman path integral formalism \cite{Nagaosa99,Wen04,Altland10,Saito20a,Saito20b,Saito20c,Saito20d}.
After eliminating $(r,\phi)$, we confirm the dispersion relationship with the opening up of the energy gap 
\begin{eqnarray}
\Delta
=\hbar \delta  w_0 (n+m+1) 
=m^{*} v_0^2,
\end{eqnarray}
where the overall shift of the energy is $\hbar \delta  w_0 = v_0 g$, $n$ is the radial quantum number, $m$ is the magnetic orbital angular momentum along the principal axis of $z$, and $m^{*}$ is the effective mass of the photon in a GRIN fibre \cite{Nagaosa99,Wen04,Altland10,Saito20a,Saito20b,Saito20c,Saito20d}.
Please note the similarity of the original Einstein theory of relativity to assign the rest mas of $m$ to its energy as $E=mc^2$ \cite{Einstein05,Einstein05b}.
The emergence of the effective mass is attributed to the broken $SU(2)$ symmetry of photons due to lasing \cite{Saito20d}.
It is interesting to note that the obtained dispersion relationship is quite similar to the Bardeen-Cooper-Schrieffer (BCS) theory of superconductivity \cite{Bardeen57,Anderson58,Bogoljubov58,Nambu59,Schrieffer71,Goldstone62,Higgs64}.
The remaining degree of freedom is the propagation of light along $z$ in a GRIN fibre, which is described by the Schr\"odinger-like equation
\begin{eqnarray}
i \hbar \partial_t \psi_z
=
-
\frac{\hbar^2}{2 m^{*}}
\Box_2 
\psi_z,
\end{eqnarray}
where $\psi_z$ is the wavefunction of a photon, $\hbar=h/(2\pi)$ is the Dirac constant, and we have re-defined the d'Alembertian
\begin{eqnarray}
\Box_2
=  
\frac{1}{v_0^2} \partial_t^2
-
\partial_z^2,
\end{eqnarray}
to account for the reduced speed of light in a material.
The wavefunction along $z$ becomes a simple plane wave,
\begin{eqnarray}
\psi_z = {\rm e}^{i k z - i \omega t},
\end{eqnarray}
while the dispersion for the guided mode \cite{Saito20d} becomes
\begin{eqnarray}
E
=
\Delta
+
\sqrt{\Delta^2+ (v_0 p)^2}.
\end{eqnarray}
In deriving this energy-momentum dispersion relationship, we assumed de-Broglie relationship
\begin{eqnarray}
E &=& \hbar \omega \\
p &=& \hbar k,
\end{eqnarray}
where $\omega$ is the angular frequency and $k=2\pi/\lambda$ is the wavenumber for the photon with the wavelength of $\lambda$.

\subsection{Lorentz transformation in a uniform material}
First, we examine the weak coupling limit of $g\rightarrow 0$.
In this case, the energy gap vanishes, $\Delta \rightarrow 0$, and the wave equation becomes
\begin{eqnarray}
\left[
  \partial_z^2-\frac{1}{v_0^2} \partial_t^2
\right]
\psi_z(t)
=0.
\end{eqnarray}
Thus, a photon is massless with the reduced velocity of $v_0$ in a uniform material of the refractive index of $n_0$.
The dispersion relationship is linear, 
\begin{eqnarray}
\omega
=
v_{p}
k
=
\frac{c}{n_0}
k,
\end{eqnarray}
and the phase velocity $v_p$ is given by $v_{p}=v_{0}=c/n_0$.

By applying the Lorentz transformation to the wave equation, we obtain the corresponding wave equation in the moving frame as 
\begin{eqnarray}
&&
\left[
  \partial_z^2-\frac{1}{v_0^2} \partial_t^2
\right]
\psi_z(t) 
\nonumber \\
=&&
\left[
\frac{1-n_0^2 \beta^2}{1-\beta^2}
\left(
  \partial_{z^{'}}^2
-\frac{1}{c^2} 
\frac{n_0^2-\beta^2}{1-n_0^2 \beta^2}
\partial_{t^{'}}^2
\right)
\right.
\nonumber \\
&&
\left.
-
2
\frac{\beta}{c}
\frac{1-n_0^2}{1-\beta^2}
\partial_{t^{'}}
\partial_{z^{'}}
\right]
\psi_{z^{'}} (t^{'})
=0.
\end{eqnarray}
Inserting the trial function of the form, 
\begin{eqnarray}
\psi_{z^{\prime}}(t^{\prime})
=
{\rm e}^{i k^{\prime} z^{\prime} - i \omega^{\prime} t^{\prime}},
\end{eqnarray}
we obtain the dispersion relationship in the moving frame,
\begin{eqnarray}
\omega^{\prime} 
=
\frac{1-\beta n_0}{1-\frac{\beta}{n_0}}
v_0
k^{\prime}.
\end{eqnarray}
The phase velocity in the frame is given by
\begin{eqnarray}
v^{\prime}_p
=\frac{\omega^{\prime}}{k^{\prime}}
=
\frac{1-\beta n_0}{n_0-\beta}
c.
\end{eqnarray}

We can obtain the same dispersion relationship, by simply inserting the Lorentz transformation into the original wavefunction as 
\begin{eqnarray}
\psi_{z^{\prime}}(t^{\prime})
&=&
{\rm e}^{ik\gamma \left( 1-\frac{\beta}{n_0} \right) z^{\prime} 
-i \omega \gamma \left( 1-\beta n_0 \right) t^{\prime}}
\\
&=&
{\rm e}^{i k^{\prime} z^{\prime} - i \omega^{\prime} t^{\prime}},
\end{eqnarray}
which leads
\begin{eqnarray}
k^{\prime}
&=&
\gamma \left( 1-\frac{\beta}{n_0} \right) k
=
\gamma \left(\frac{n_0-\beta}{n_0} \right) k
\\
\omega^{\prime} 
&=&
\gamma \left( 1-\beta n_0 \right) 
\omega .
\end{eqnarray}

\subsubsection{Vacuum limit}
We check the obtained dispersion in the know limit of the vacuum, such that we take the limit of $n_0 \rightarrow 1$.
In this case, we reproduce a standard theory of relativity \cite{Einstein05,Einstein05b,Jackson99}.
We obtain
\begin{eqnarray}
\omega^{'}
\rightarrow
\gamma (1-\beta) \omega
&=&
\sqrt{
\frac{1-\beta}{1+\beta}
}
\omega
\\
k^{'}
\rightarrow
\gamma (1-\beta) 
k
&=&
\sqrt{
\frac{1-\beta}{1+\beta}
}
k,
\end{eqnarray}
from which we obtain the Doppler effect for the light by setting $k^{\prime}=2\pi/\lambda^{\prime}$, we obtain
\begin{eqnarray}
\lambda^{'}
=
\sqrt{
\frac{1-\beta}{1+\beta}
}
\ 
\lambda.
\end{eqnarray}
We expect the red-shift for $v_z>0$, since the light source is relatively going away such that the wavelength is elongated for the observer moving away from the light source.
On the other hand, the blue-shift is expected for $v_z<0$, since the light source is approaching to the observer.
We can also confirm the principle of relativity by exchanging $\lambda \leftrightarrow \lambda^{\prime}$ and $\beta \leftrightarrow -\beta$ at the same time, the relationship between $\lambda$ and $\lambda^{\prime}$ are not altered.

\subsubsection{Uniform material}
We assumed the material is at rest in the frame of $(t,x,y,z)$ and the moving frame of $(t^{\prime},x^{\prime},y^{\prime},z^{\prime})$ is not equivalent to the original frame any more.
In this case, the wavenumber of $k^{\prime}$ is always positive, while the angular frequency of $\omega^{\prime}$ can change its sign upon increasing the frame velocity of $\beta$.
Consequently, the phase velocity of $v_p^{\prime}$ can also change sign, such that the light is seen to be propagating to the backward, if the frame is moving faster than the speed of light in a material, which is indeed possible as is know for the case of the Cherenkov radiation \cite{Cherenkov34,Frank37,Cherenkov37,Cherenkov37b,Cherenkov37c,Cherenkov86,Baryshevsky17}.
We can see the consequence of the Lorentz transformation by examining several typical limits of the obtained phase velocity of $v_p^{\prime}$ (Table \ref{Table-I}).
The details of the calculated parameters are discussed in the next section (Fig. 1).

 \begin{table}[h]
\caption{\label{Table-I}
Phase velocity of light observed from a moving frame in a uniform material.}
\begin{ruledtabular}
\begin{tabular}{lll}
Limit&Phase velocity & Comment\\
\colrule
$n_0 \rightarrow 1$ & $v_p^{\prime}$ & Vacuum limit \\
$\beta \rightarrow 0$ & $v_p^{\prime} \rightarrow v_p$ & Rest limit \\
$0<v_z<v_p$ & $0<v_p^{\prime}<v_p$ & Frame moving slower than light \\
$v_z \rightarrow v_p$ & $v_p^{\prime} \rightarrow 0$ & Stopping light \\
$v_p < v_z$ & $v_p^{\prime} <0$ &  Frame moving faster than light \\
$\beta \rightarrow 1$ & $v_p^{\prime} \rightarrow -c$ & Maximum velocity
\end{tabular}
\end{ruledtabular}
\end{table}

\subsection{Lorentz transformation in a GRIN fibre}
In a GRIN fibre, the dispersion relationship is different from a uniform material, due to the band-gap opening, as we have outlined above \cite{Allen92,Yariv97,Saito20a,Saito20b,Saito20c,Saito20d}.
We can obtain the corresponding dispersion relationship observed in a moving frame.
The main assumption is the plane wave form of the solution along $z$ and $t$ and the validity of the space-time relationship by the Lorentz transformation.
Consequently, we obtain
\begin{eqnarray}
\psi_{z^{\prime}}(t^{\prime})
&=
{\rm e}^{ik\gamma \left( \beta c t^{\prime} + z^{\prime} \right)  
-i \omega \gamma \left( t^{\prime}+ \frac{\beta}{c} z^{\prime} \right) }
\equiv
{\rm e}^{i k^{\prime} z^{\prime} - i \omega^{\prime} t^{\prime}},
\end{eqnarray}
which gives
\begin{eqnarray}
\omega^{\prime} 
&=&
\gamma \left( \omega - \beta c k \right) 
\\
k^{\prime}
&=&
\gamma \left( k- \beta \frac{\omega}{c} \right) .
\end{eqnarray}
The relationship is simply summarised as the Lorentz transformation of 
\begin{eqnarray}
\left (
  \begin{array}{cc}
   \omega^{'}
\\
   ck^{'}
  \end{array}
\right)
=
L
\left (
  \begin{array}{cc}
   \omega
\\
   ck
  \end{array}
\right).
\end{eqnarray}
This is equivalent to impose the de-Bloglie relationship in the moving frame as 
\begin{eqnarray}
E^{\prime} &=&\hbar \omega^{\prime} \\
p^{\prime} &=& \hbar k^{\prime},
\end{eqnarray}
which gives the Lorentz transformation of the energy-momentum relationship as 
\begin{eqnarray}
\left (
  \begin{array}{cc}
   E^{'}
\\
   p^{'}
  \end{array}
\right)
=
L
\left (
  \begin{array}{cc}
   E
\\
   p
  \end{array}
\right),
\end{eqnarray}
which must be valid for an arbitrary dispersion $E=E(p)$ beyond the dispersion for a GRIN fibre.
The relative symmetry against a frame exchange does not exist any more with a material, because the rest frame with a material is different from the moving frame.
Consequently, we obtain non-trivial Doppler effects, as shown in the next section.

\section{Results}

\subsection{Doppler effects in a uniform material}

\begin{figure}[h]
\begin{center}
\includegraphics[width=8.5cm]{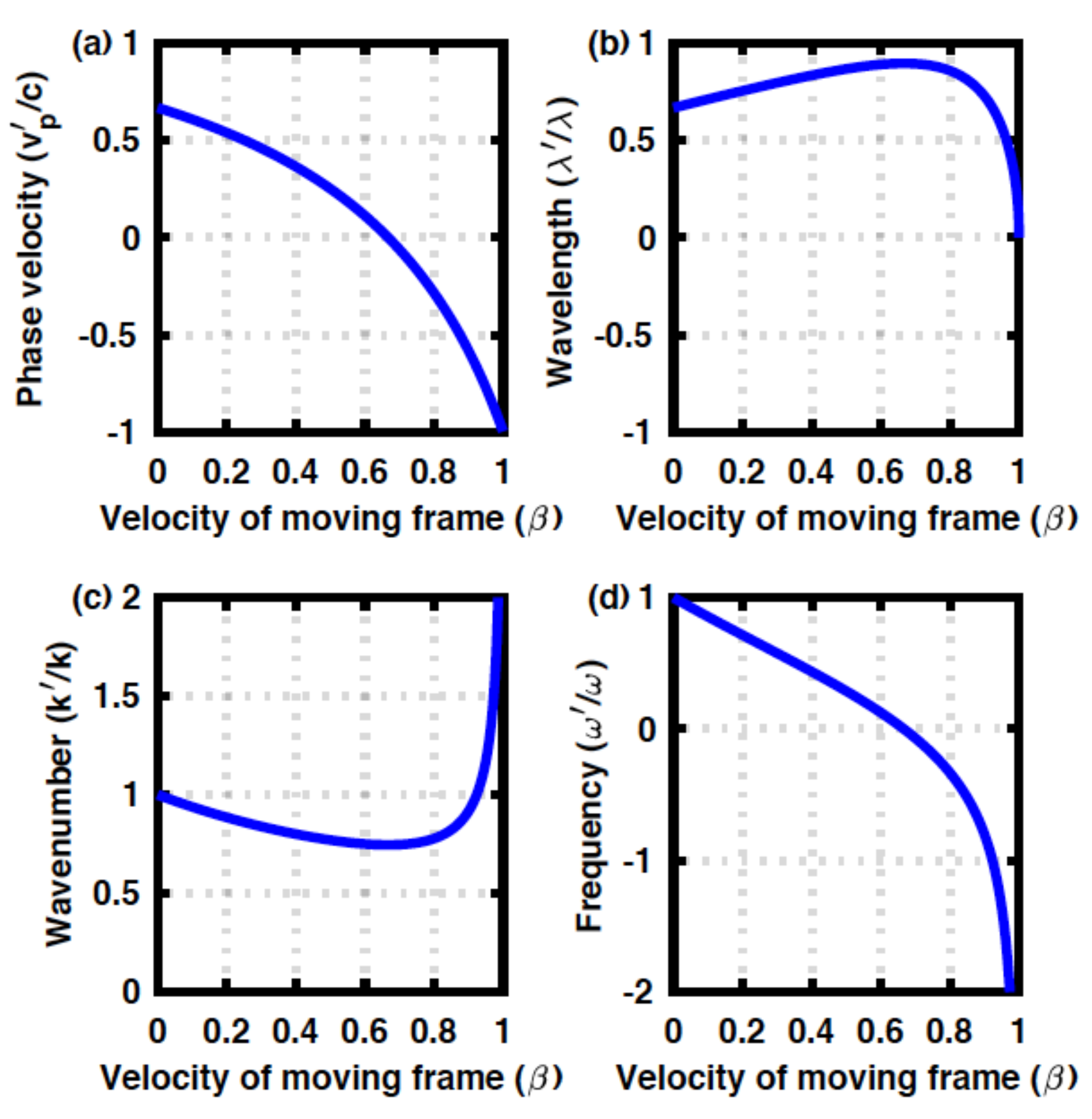}
\caption{
Doppler effects in a uniform material observed from a moving frame.
(a) Phase velocity changes sign at $v_z=v_p$.
(b) Wavelenth is red-shifted for $v_z<v_p$, while it starts to exhibit blue-shift for $v_p<v_z$.
(c) Wavenumber is always positive, while (d) frequency changes sign.
}
\end{center}
\end{figure}

First, we show numerical results in a uniform material (Fig. 1).
We assumed $n_0=1.5$ and $\Delta=0$, and the wavelength of $\lambda=1.5$ ${\rm \mu}$m, considering typical parameters of a glass fibre \cite{Yariv97}.
Upon increasing the frame velocity of $\beta$, the phase velocity of $v_p^{\prime}$ decreases and changes sign at $v_z=v_p$ and $v_p^{\prime}$ approaches $-c$ in the maximum limit of $\beta \rightarrow 1$ (Fig. 1 (a)).

We expect the red-shift for $v_z<v_p$, since the observer in the moving frame is going away from the light source in the rest frame, such that the wavelength is elongated, while the blue-shift is expected for $v_p<v_z$ (Fig. 1 (b)).
By assuming $k^{\prime}=2\pi/\lambda^{\prime}$, we obtain the wavelength in the moving frame
\begin{eqnarray}
\lambda^{\prime}
=
\frac
{\sqrt{1-\beta^2}}
{n_0 - \beta}
\lambda.
\end{eqnarray}
We confirm the appropriate limit of $\lambda^{\prime}\rightarrow \lambda$ for $\beta\rightarrow 0$, while the limit of $\lambda^{\prime}\rightarrow 0$ for $\beta\rightarrow 1$ might be non-trivial.
We expect the peak of the wavelength at 
\begin{eqnarray}
\frac{\partial \lambda^{\prime}}{\partial \beta}
=
\frac{1-\beta n_0}
{(n_0 - \beta)^2 \sqrt{1-\beta^2}}
=0,
\end{eqnarray}
which indeed gives $v_z=v_p$.
We are considering the continuous wave, emitted from the light source in the rest frame, rather than a pulsed operation. 
If the frame is moving above the speed of light in the rest frame, the frame is approaching to the light, which was emitted earlier, such that the wavelength of the light is observed shorter than that in the rest frame.
The wavenumber is always positive (Fig. 1 (c)), since the refractive index of a material is always larger than unity ($n_0>1$) and the frame cannot move larger than $c$ ($\beta<1$).

On the other hand, the angular frequency of $\omega^{\prime}$ changes sign at $v_z=v_p$, such that the polarisation state starts to rotate in the opposite way, as if the time is going backward (Fig. 1 (d)).
This could be considered by defining a chiral operator defined by
\begin{eqnarray}
\hat{\chi}_z
&=&{\rm sgn}\left( \hat{v}_p \right)
={\rm sgn}\left( \frac{\hat{\omega}}{\hat{k}} \right)\\
&=&-{\rm sgn}\left( \frac{\partial_t}{\partial_z} \right).
\end{eqnarray}
In the moving frame, we consider
\begin{eqnarray}
\chi^{\prime}
={\rm sgn}\left( v_p^{\prime} \right)
={\rm sgn}\left( \frac{\omega^{\prime}}{k^{\prime}} \right),
\end{eqnarray}
which changes sign at $v_z=v_p$.
Therefore, the helicity is reversed upon increasing $\beta$.

\subsection{Doppler effects in a GRIN fibre}

\begin{figure}[h]
\begin{center}
\includegraphics[width=8.5cm]{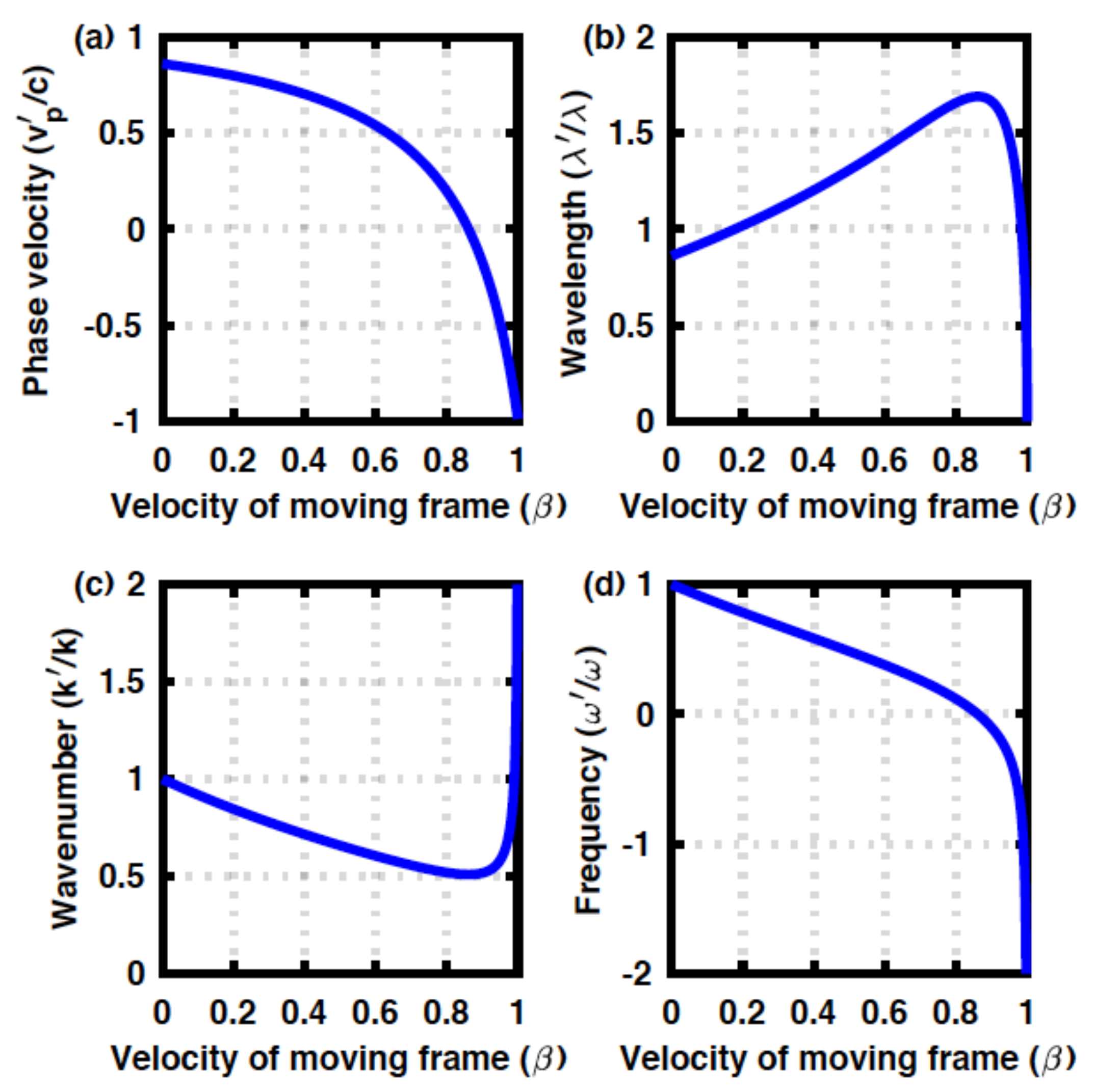}
\caption{
Doppler effects in a GRIN fibre at $\Delta=0.2\hbar \omega$.
(a) Phase velocity, (b) wavelength, (c), wavenumber, and (d) frequency.
}
\end{center}
\end{figure}

Next, we consider the Doppler effects in a GRIN fibre.
We assume the same core of $n_0$ for the wavelength of $\lambda=1.5$ ${\rm \mu}$m, while the energy gap of $\Delta$ is chosen as a parameter.

The numerical results at $\Delta=0.2 \hbar \omega$ are shown in Fig. 2. 
The qualitative features are not changed for the case of a uniform material (Fig. 1).
The critical frame velocity, required to change the sing of the phase velocity, is increased due to the opening of the band gap (Fig. 2 (a)).
We expect more significant red- and blue-shifts upon increasing $\beta$ (Fig. 2 (b)), but $k^{\prime}$ is always positive and $\omega^{\prime}$ changes sign, as before.

\begin{figure}[h]
\begin{center}
\includegraphics[width=8.5cm]{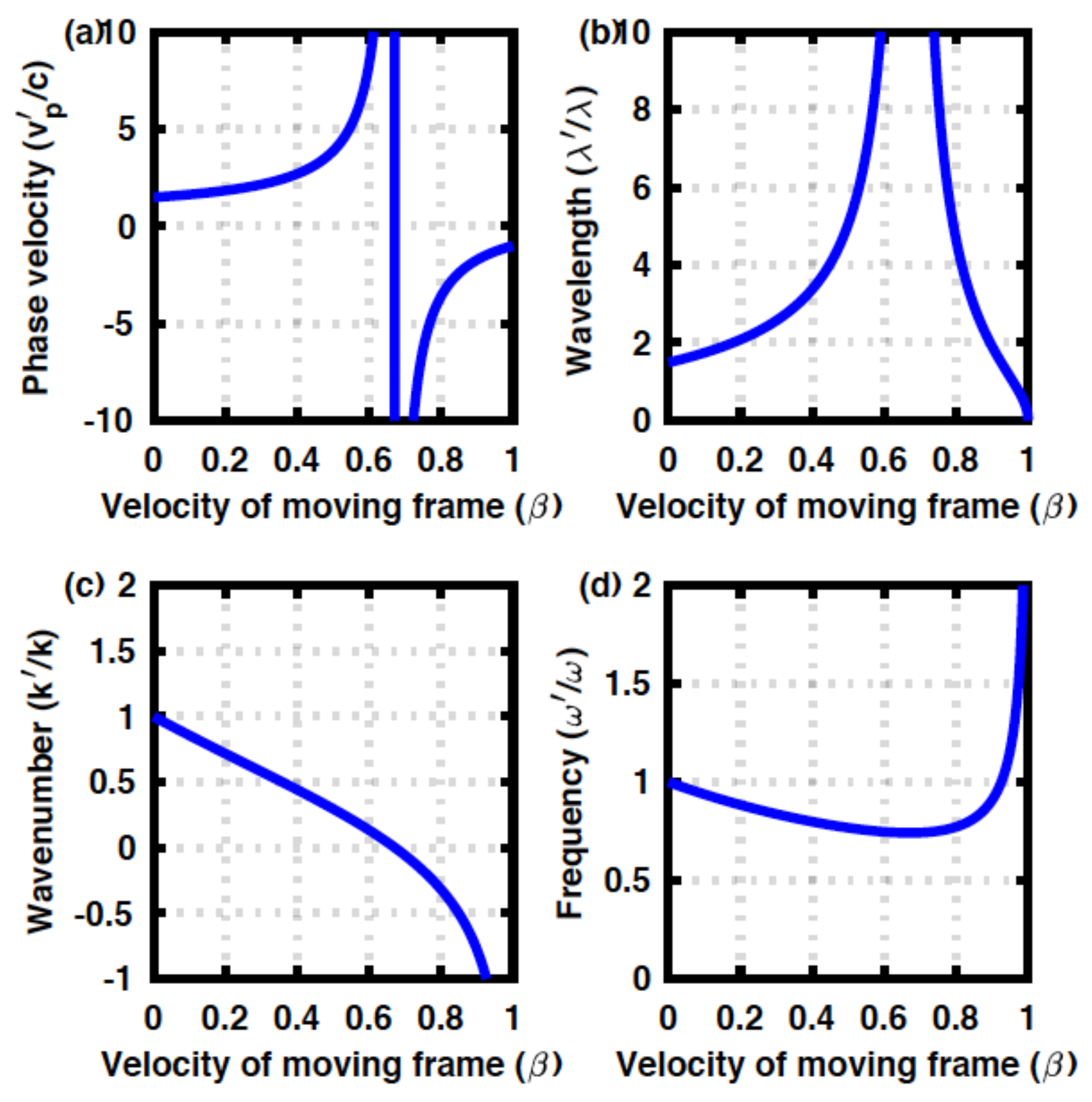}
\caption{
Doppler effects in a GRIN fibre at $\Delta=0.4\hbar \omega$.
(a) Phase velocity diverges and it could be larger than $c$. 
(b) Wavelength also diverges. 
(c) Wavenumber changes sign, while (d) frequency is always positive.
}
\end{center}
\end{figure}

On the other hand, for the larger $\Delta$ at $0.4 \hbar \omega$, Doppler effects are even more anomalous. 
In this case, $k^{\prime}$ changes sign, while $\omega^{\prime}$ is always positive.
This is attributed to the larger contributions to the total energy of $\hbar \omega$ from orbital degrees of freedoms through the radial oscillations and/or photonic orbital angular momentum, characterised by $n$ and $m$ \cite{Allen92,Yariv97,Saito20a,Saito20b,Saito20c,Saito20d}.
As a result, the contribution of the kinetic energy for the propagation along $z$ is limited, such that the frame is easier to go beyond the speed of the light, which allows $k^{\prime}$ to change the sign (Fig. 3  (a)).
For $k^{\prime}<0$, we have assumed $k^{\prime}=-2\pi/\lambda^{\prime}$ to extract the wavelength of $\lambda^{\prime}$ in the moving frame (Fig. 3 (b)). 

\begin{figure}[h]
\begin{center}
\includegraphics[width=8.5cm]{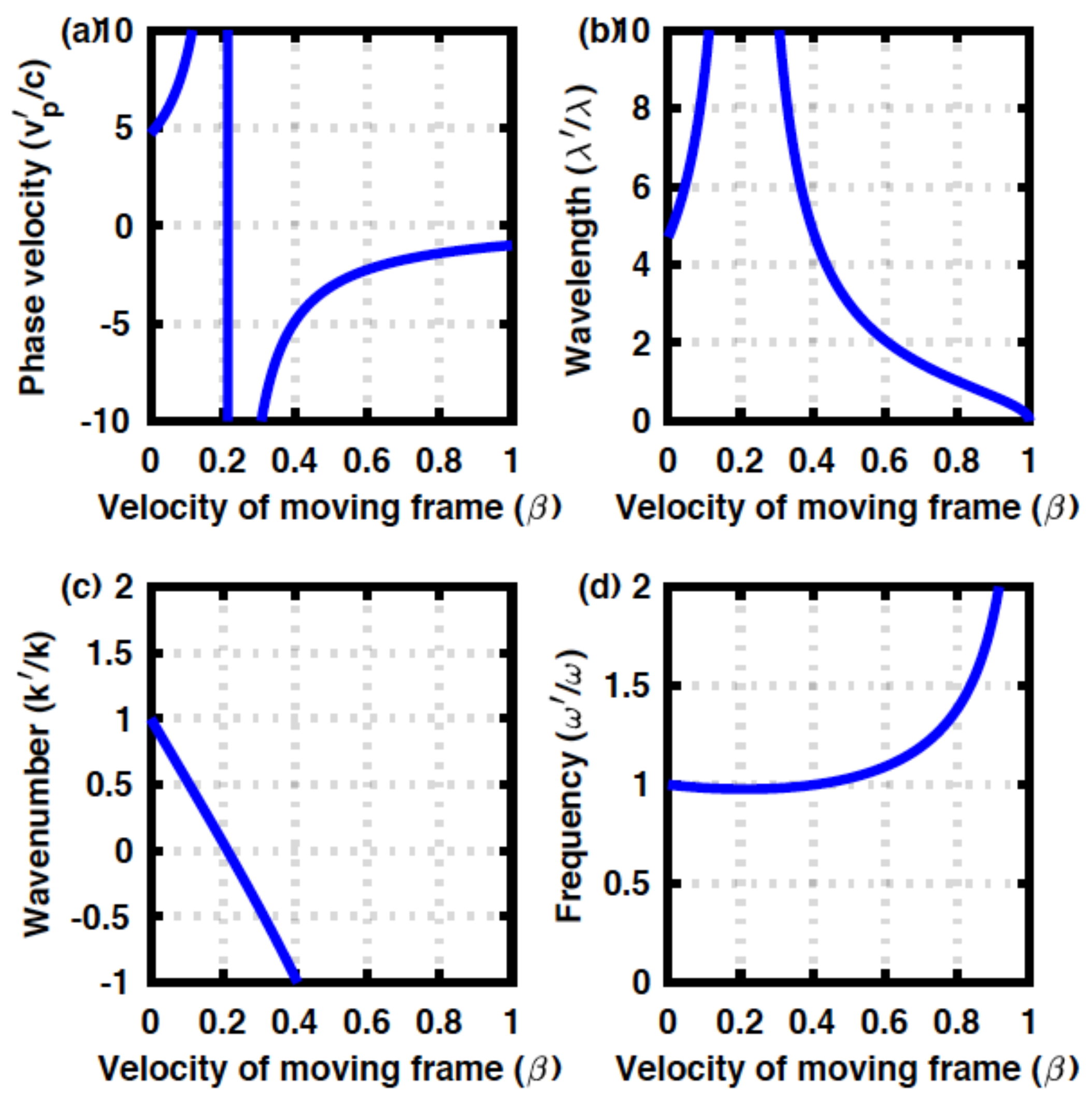}
\caption{
Doppler effects in a GRIN fibre at $\Delta=0.49\hbar \omega$.
(a) Phase velocity, (b) wavelength, (c) wavenumber, and (d) frequency.
The contributions from orbital degrees of freedom are dominated over the kinetic energy along the fibre.
}
\end{center}
\end{figure}

The maximum energy gap is $\Delta=0.5\hbar \omega$, where we cannot expect any propagation along $z$, and the optical mode is trapped solely in the direction perpendicular to the fibre.
Close to this limit, we assumed $\Delta=0.49\hbar \omega$ and the results are shown in Fig. 4. 
$v_p^{\prime}$ changes its sign even at the smaller $\beta$, as expected for the limited kinetic energy.

\begin{figure}[h]
\begin{center}
\includegraphics[width=6cm]{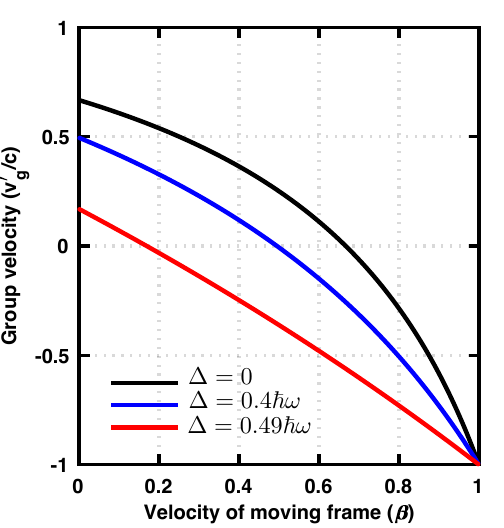}
\caption{
Group velocity of light in a GRIN fibre for $\Delta=0$ (uniform material without confinement), $\Delta=0.4 \hbar \omega$, and  $\Delta=0.49 \hbar \omega$ (strong confinement in the core).
In all cases, the group velocity is less than the speed of light in a vacuum, satisfying the causality.
}
\end{center}
\end{figure}

Regardless of the anomalous phase velocity of $|v_p^{\prime}|$, exceeding to $c$ (Figs. 3(a) and 4(a)), this does not mean the violation of the relativity at all, because the optical communication is determined by the group velocity, defined by
\begin{eqnarray}
v_{\rm g}
&=&
\frac{d k}{d \omega}\\
v_{\rm g}^{\prime}
&=&
\frac{d k^{\prime}}{d \omega^{\prime}}.
\end{eqnarray}
As shown in Fig. 5, $|v_{\rm g}^{\prime}|$ is always smaller than $c$, such that the optical communication beyond $c$ is strictly prohibited.
The critical velocity of $\beta$ to change the sign of $v_{\rm g}^{\prime}$ does not necessarily coincide with the velocity of $\beta$ to change the sign of $v_{\rm p}^{\prime}$.

\subsection{Impacts on polarisation states}
Finally, we discuss the implications of our considerations for understanding of the polarisation states of photons.
Before discussing the application of the general theory of relativity to the GRIN fibre, we need to clarify the definition of the polarisation states, because the direction of the apparent propagation changes in the frame moving faster than the phase velocity of the light in the rest frame.

\subsubsection{Lights propagating in opposite directions}
\begin{figure}[h]
\begin{center}
\includegraphics[width=8.5cm]{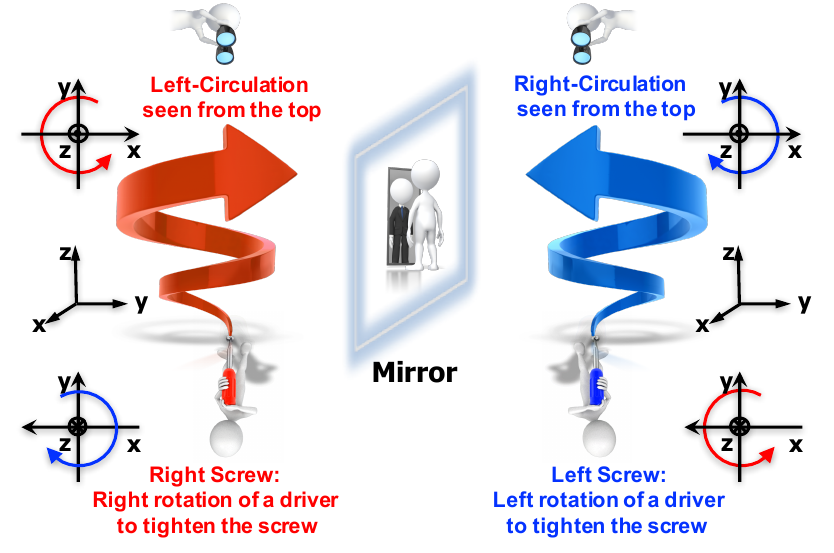}
\caption{
Impacts of a mirror.
Polarisation states are defined by the motion of the phase front, seen from the top of the light at the detector side.
The left-circular-polarised state is characterised by the anti-clock-wise rotation, while the right-circular-polarised state exhibit the clock-wise rotation.
A mirror exchanges these circularly polarised states to the opposite of each other.
Some researchers prefer to define by the motion of the phase front, seen from the source side similar to the definitions of right and left screws.
We will not take the latter notation, to follow the mathematical notation to see the right-handed $(x,y,z)$ coordinate to see from the top of the $+z$ axis.
}
\end{center}
\end{figure}

We clarify the definition of the polarisation states \cite{Saito20a}, in particular, the direction of the rotation (Fig. 6).
There are a lot of different choices of the conventions \cite{Jackson99,Yariv97,Goldstein11,Gil16,Saito20a}, and any notation is acceptable as far as it is used consistently.
We prefer to define the polarisation state, seen from the detector side, because it is straightforward to describe the motion of the phase front in a standard right-handed $(x,y,z)$ coordinate (Fig. 6).
We assume that the plane wave of the form ${\rm e}^{ikz-i\omega t}$ is propagating along the $+z$ direction (Fig. 7(a)), and the principal axis ($S_3$) of the polarisation state is locked along the direction of the propagation \cite{Saito20a}.
In our definition, the left-circular-polarised state ($|{\rm L} \rangle$) is located at $S_3=+1$ in the normalised Poincar\'e sphere, while the right-circular-polarised state ($|{\rm R} \rangle$) is located at $S_3=-1$ (Fig. 7(e)).
These states are mirror images of each other, and in fact, a mirror can change the circular polarised states to the opposite ones \cite{Yariv97,Saito20a,Goldstein11,Gil16}.
 
\begin{figure}[h]
\begin{center}
\includegraphics[width=7.0cm]{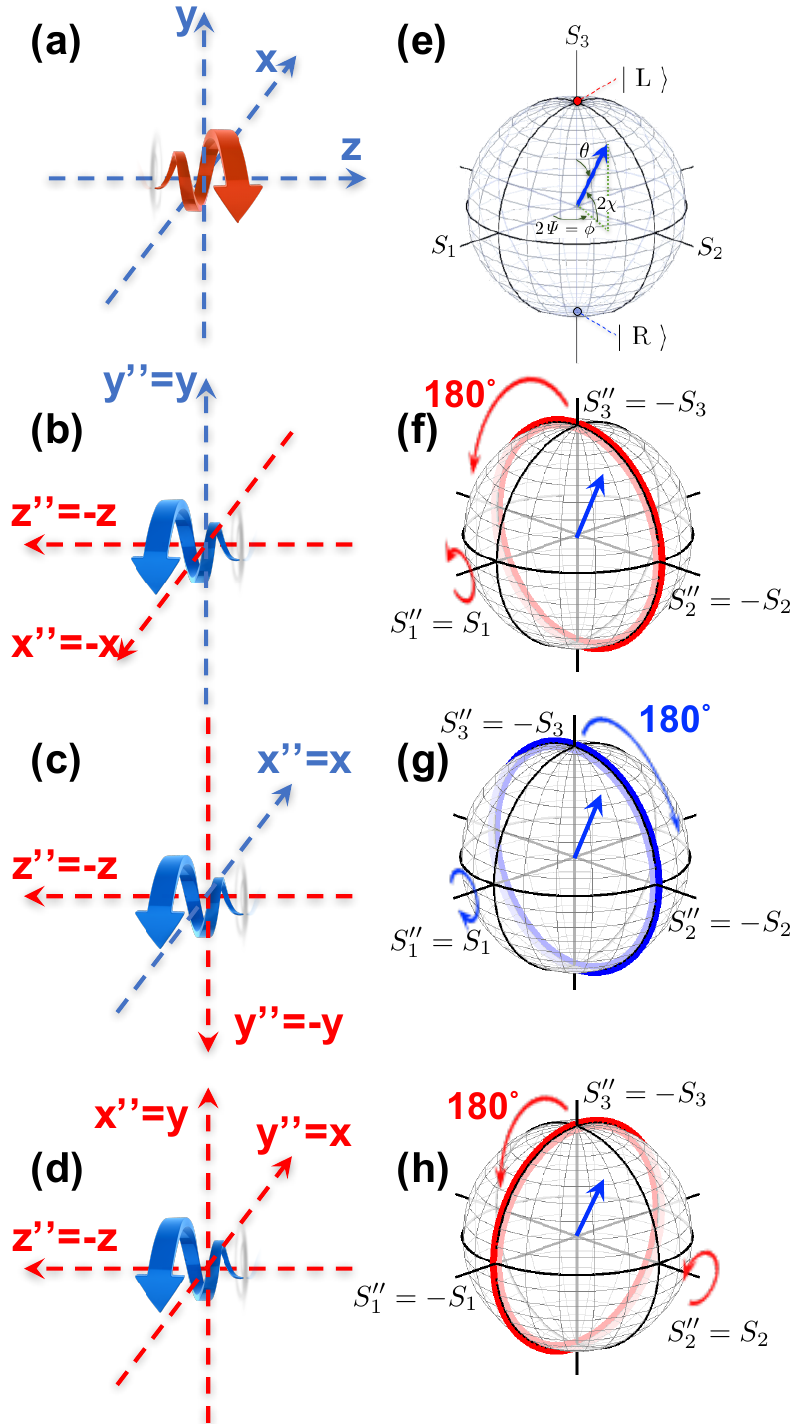}
\caption{
Choices of frames and polarisation states.
(a) The original frame of $(x,y,z)$ for the light propagating along the $+z$ direction, and (e) the corresponding Poinca\'e sphere.
(b)-(d) The frames of $(x^{\prime \prime},y^{\prime \prime},z^{\prime \prime})$ for the light propagating along the $z^{\prime \prime}=-z$ direction, and (f)-(h) the corresponding Poinca\'e spheres and their relevance to the original Poincar\'e sphere.
Here, all frames are at rest against others.
(b) $(x^{\prime \prime},y^{\prime \prime},z^{\prime \prime})$ was made by the rotation along $y$, which is equivalent to rotate $\pi$ along $S_1$ (f).
(c) $(x^{\prime \prime},y^{\prime \prime},z^{\prime \prime})$ was made by the rotation along $x$, which is equivalent to rotate $-\pi$ along $S_1$ (g).
(d) $(x^{\prime \prime},y^{\prime \prime},z^{\prime \prime})$ was made by the subsequent rotation $\pi$ from (b), or equivalently by the rotation of $-\pi$ from (c), which is equivalent to rotate $\pi$ along $S_2$ (h).
}
\end{center}
\end{figure}

We consider how we should describe the polarisation states for the light propagating to the opposite direction (Fig. 7), since we encountered the situation that the moving frame can go faster than the speed of the light in a material, for which the light is seen to go backward, in the previous sections.
The situation is similar to consider the reflection from the mirror (Fig. 6), since a mirror changes the direction of propagation as well as the polarisation state.
For example, we consider the left-circular-polarised light propagating along $+z$ direction (Fig. 7 (a)), which is characterised by $S_3=+1$ (Fig. 7 (e).
We consider this light is reflected backwards without changing the rotation of the phase front (Fig. 7 (b)), while the direction of the propagation is opposite ($-z$).
If we are keeping observing the phase front, seen from the $+z$ direction, the circulation is unaffected as the anti-clock-wise rotation.
However, we defined that the polarisation state must be identified from the detector side, which is $-z$ direction (Fig. 7 (b)).
Thus, we define a new frame of $(x^{\prime \prime},y^{\prime \prime},z^{\prime \prime})$, assuming $z^{\prime \prime}=-z$ to clarify the polarisation state.
The reflected light propagating along $z^{\prime \prime}=-z$ is now circulating to the clock-wise direction, seen from the $+z^{\prime \prime}$ direction, thus it should be described by the right-circular-polarised state with $S_3^{\prime \prime}=-1$ (Fig. 7 (f)).
Considering the opposite direction of the propagation, it is natural to assign $S_3^{\prime \prime}=-S_3$, if we would like to keep using the original axes for Stokes parameters.
Here, we still have a freedom to choose the relative phase of axes (Figs. 7 (b)-(d)) to the original one (Fig. 7 (a)).
These choices correspond to how to rotate the original Poincar\'e sphere at the angle of $\pi$ (Figs. 7 (f)-(h)).
The rotation of the polarisation state is described by a rotation operator in $SU(2)$ Lie algebra \cite{Jones41,Payne52,Baym69,Sakurai14,Jackson99,Yariv97,Goldstein11,Gil16}.
If we use the chiral LR-basis, the rotation operator becomes

\begin{eqnarray}
\mathcal{D}_{\rm LR}({\bf n},{\it \Delta \delta})
&=&
\exp 
\left (
  -\frac{i {\bm \sigma}\cdot {\bf n} {\it \Delta \delta}}{2}
\right)
\\
&=&
{\bf 1}
\cos 
\left(
  \frac{{\it \Delta \delta}}{2}
\right)
-
i {\bm \sigma}\cdot {\bf n}
\sin 
\left(
  \frac{{\it \Delta \delta}}{2}
\right),
\end{eqnarray}
where ${\bf n}$ is the unit vector for the rotational axis, ${\bm \sigma}=(\sigma_1,\sigma_2,\sigma_3)$ are Pauli matrices, and ${\it \Delta \delta}$ is the angle of the rotation.
For example, the rotation along $S_1$ for $\pi$ (Fig. 7 (f)) is given by
\begin{eqnarray}
\mathcal{D}_{\rm LR}(S_1,\pi)
=-i\sigma_{1},
\end{eqnarray}
while the opposite rotation for Fig. 7 (g) is described by 
\begin{eqnarray}
\mathcal{D}_{\rm LR}(S_1,-\pi)
=i\sigma_{1}.
\end{eqnarray}
Similarly, the rotation along $S_2$ for $\pi$ (Fig. 7 (h)) is given by 
\begin{eqnarray}
\mathcal{D}_{\rm LR}(S_2,\pi)
=-i\sigma_{2}.
\end{eqnarray}
These rotations are connected each other.
For example, the coordinate of Fig. 7(d) is realised by rotating Fig. 7 (b) for $\pi/2$, which correspond to the $\pi$ rotation along $S_3$ \cite{Saito20b},
\begin{eqnarray}
\mathcal{D}_{\rm LR}(S_3,\pi)
=-i\sigma_{3}.
\end{eqnarray}
In fact, we confirm
\begin{eqnarray}
-i\sigma_{3}\left( -i\sigma_{1} \right)
=
-\sigma_{3}\sigma_{1}
=-i\sigma_{2}.
\end{eqnarray}
Similarly, we can rotate the coordinate of Fig. 7 (c) for $\pi/2$, which correspond to the $-\pi$ rotation along $S_3$, and we confirm
\begin{eqnarray}
+i\sigma_{3}\left( i\sigma_{1} \right)
=
-\sigma_{3}\sigma_{1}
=-i\sigma_{2}.
\end{eqnarray}
The arbitrary degree of freedom to chose the $(x^{\prime \prime},y^{\prime \prime})$ axes is actually not restricted to the reflected beam.
For example, if we have a linear diagonally polarised state, which is described by $S_2=1$, by changing the definition of the $x$-axis by rotating $45^{\circ}$, it can also be regarded as the horizontally polarised state of $S_1=1$. Therefore, the difference of the apparent polarisation states between $S_1$ and $S_2$ simply depends on the choice of the frame.

Among various arbitrary choices of the frame for the reflected light (Figs. 7 (b)-(d)), however, one of the most sensible choice would be that of Fig. 7 (d).
In this case, the impact of the frame exchange is similar to the unitary transformation by the mirror operation of 
\begin{eqnarray}
M_{\rm LR}
=
-i
\left (
  \begin{array}{cc}
    0 & -i \\
    i & 0
  \end{array}
\right)
=
-i\sigma_2,
\end{eqnarray}
whose impact on the spin operators would be
\begin{eqnarray}
M_{\rm LR}^{\dagger}
\sigma_1
M_{\rm LR}
&=&
(+i\sigma_2)
\sigma_1
(-i\sigma_2)
=
\sigma_2\sigma_1\sigma_2
=-\sigma_1
=-\sigma_1^{*}
\nonumber \\
\\
M_{\rm LR}^{\dagger}
\sigma_2
M_{\rm LR}
&=&
(+i\sigma_2)
\sigma_2
(-i\sigma_2)
=
\sigma_2\sigma_2\sigma_2
=+\sigma_2
=-\sigma_2^{*}
\nonumber \\
\\
M_{\rm LR}^{\dagger}
\sigma_3
M_{\rm LR}
&=&
(+i\sigma_2)
\sigma_3
(-i\sigma_2)
=
\sigma_2\sigma_3\sigma_2
=-\sigma_3
=-\sigma_3^{*},
\nonumber \\
\end{eqnarray}
where $^{*}$ is the complex conjugate and $^{\dagger}$ is the Hermite conjugate, which involves the transpose of the matrix in addition to the complex conjugate.
In this convention, we understand that the polarisation state for the light propagating in the opposite direction of $-z$ is described by the complex conjugate representation of the Lie algebra \cite{Georgi99,Pfeifer03} in the original frame as
\begin{eqnarray}
\overline{\bm \sigma}
=
\left (
  \begin{array}{c}
    \overline{\sigma}_1\\
    \overline{\sigma}_2\\
    \overline{\sigma}_3
  \end{array}
\right)
=
-
\left (
  \begin{array}{c}
    \sigma_1^{*}\\
    \sigma_2^{*}\\
    \sigma_3^{*}
  \end{array}
\right)
=
\left (
  \begin{array}{c}
    -\sigma_1\\
   + \sigma_2\\
   - \sigma_3
  \end{array}
\right),
\end{eqnarray}
which is consistent with Fig. 7 (h).
The complex conjugate representation also satisfies the same commutation and anti-commutation relationships with those of the original Pauli matrices \cite{Georgi99,Pfeifer03}, such that we confirmed the duality of representations.
Therefore, if we would like to keep working in the original frame of $(x,y,z)$ for the light propagating in the opposite direction, we should use the complex conjugate of the spin operators, defined by
\begin{eqnarray}
\overline{S}_{x}
&=&
-
\hbar 
     \bm{\psi}_{\rm LR}^{\dagger}
\sigma_1
\bm{\psi}_{\rm LR}
=-{S}_{x},
\\
\overline{S}_{y}
&=&
+
\hbar 
     \bm{\psi}_{\rm LR}^{\dagger}
\sigma_2
\bm{\psi}_{\rm LR}
=+{S}_{y},
\\
\overline{S}_{z}
&=&
-
\hbar 
     \bm{\psi}_{\rm LR}^{\dagger}
\sigma_3
\bm{\psi}_{\rm LR}
=-{S}_{z},
\end{eqnarray}
where the spinor representation of the creation and annihilation field operators are 
\begin{eqnarray}
\bm{\psi}_{\rm LR}^{\dagger}
&=&
(a_{\rm L}^{\dagger}, a_{\rm R}^{\dagger})
\\
\bm{\psi}_{\rm LR}
&=&
\left(
  \begin{array}{c}
     a_{\rm L} \\
     a_{\rm R}
  \end{array}
\right),
\end{eqnarray}
using the creation and annihilation operators of $a_{\rm \sigma}^{\dagger}$ and $a_{\rm \sigma}$ for photons in the polarisation states of left- ($\sigma={\rm L}$) and right- ($\sigma={\rm R}$) polarised states, respectively.

\subsubsection{Polarisation states observed from a moving frame}
Now, we are ready to discuss the polarisation state of light, seen from an observer in the frame of $(x^{\prime},y^{\prime},z^{\prime})$ moving as fast as the phase velocity of light in the rest frame of $(x,y,z)$, where the fibre optic material is placed.
We consider that the frame of $(x^{\prime \prime}=y^{\prime},y^{\prime \prime}=x^{\prime},z^{\prime \prime}=-z^{\prime})$ is at rest against the frame of $(x^{\prime},y^{\prime},z^{\prime})$.

First, we consider the weak coupling limit of $\Delta \rightarrow 0$, which corresponds to a uniform material with the refractive index of $n_0$.
The wavenumber of $k^{\prime}$ is always positive, such that the momentum of $p^{\prime}=\hbar k^{\prime}$ is always pointing towards the positive $+z^{\prime}$ direction (Fig. 1(c)).
On the other hand, $\omega^{\prime}$ changes its sign as $\beta$ is increased (Fig. 1(d)), leading to the change of the sign in $v_p^{\prime}$ for $v_p<v_z$.
Suppose that the light at the left-circularly-polarised state ($S_3=1$) in the original frame is propagating along $+z$.
As far as the velocity of the frame is small, $v_z<v_p$, the polarisation state, seen from the frame of $(x^{\prime},y^{\prime},z^{\prime})$ is not affected, and we obtain $S_3^{\prime}=1$ and the light is seen to be rotating in the anti-clock-wise direction, seen from $+z^{\prime}$.
At $v_p<v_z$, the light is seen to be propagating along $-z^{\prime}$, such that the polarisation state should be examined from the frame of $(x^{\prime \prime},y^{\prime \prime},z^{\prime \prime})$, where the light is propagating along the positive $+z^{\prime \prime}$ direction.
The rotation of the phase front, seen from $+z^{\prime \prime}$ direction, is anti-clock-wise due to the negative $\omega^{\prime}<0$ and the observation from the opposite side from the original frame of $(x,y,z)$.
Thus, we conclude $S_3^{\prime \prime}=1$, such that the light is still in the left-circular-polarised state.
If we consider the complex conjugate relationship between the frames of $(x^{\prime},y^{\prime},z^{\prime})$ and $(x^{\prime \prime},y^{\prime \prime},z^{\prime \prime})$, we obtain $S_3^{\prime }=-S_3^{\prime \prime}=-1$, such that the apparent spin expectation value of $S^{\prime}$ depends on the relative velocity of the frame, as is similar to frame-dependent momentum ($p^{\prime}$) and energy ($E^{\prime}$)  governed by Lorentz transformation.
For the linearly polarised states, we do not have to be careful too much on the direction of the oscillations, changed by the sign of $\omega^{\prime}$, because the rotation of the phase front is not involved.
However, the description of the polarisation state depends on the choice of the frame (Fig. 7 (b)-(d)).
Assume that we have chosen our preferential frame of Fig. 7 (d) and we consider the linear-horizontally-polarised state of $S_1=1$ in the original frame of $(x,y,z)$.
For $v_z<v_p$, the polarisation state is not affected, such that we expect $S_1^{\prime}=1$, while for $v_p<v_z$, we should use the frame of $(x^{\prime \prime},y^{\prime \prime},z^{\prime \prime})$ and the direction of oscillation is considered to be $y^{\prime \prime}=x^{\prime}$.
Therefore, we conclude that the polarisation state becomes $S_1^{\prime \prime}=-1$, which is vertically polarised state.
This is merely coming from the choice of the frame, and if we convert it to the frame of $(x^{\prime},y^{\prime},z^{\prime})$, we obtain $S_1^{\prime}=-S_1^{\prime \prime}=1$, which has not been changed upon increasing $\beta$.
We can consider a more complicated polarisation state, but the argument is straightforward.

Next, we consider the polarisation state in a GRIN fibre.
As far as the confinement is weak (fig. 2), the qualitative situation is the same as that for a uniform material, discussed above.
Therefore, we focus on the strong coupling limit (Figs. 3 and 4), where $k^{\prime}$ changes the sign upon increasing $\beta$ (Figs. 3(c) and 4(c)), while $\omega^{\prime}$ is always positive (Figs. 3(d) and 4(d)).
In these cases, the direction of the rotation of the polarisation state will not be changed by $\omega^{\prime}$, while we must judge the polarisation state seen from the direction of the propagation, which is changed.
Suppose we are considering the light of left-circularly-polarised state propagating $+z$ direction in the frame of $(x,y,z)$, such that the original state is $S_3=1$.
In the frame of $(x^{\prime},y^{\prime},z^{\prime})$, the direction of the propagation could be changed for $v_p^{\prime}<0$, and the polarisation state is examined from $z^{\prime \prime}$.
In this case, the phase front is seen to be rotating along the clock-wise-direction, because it is observed from the opposite side of the original frame of $z$ and $\omega^{\prime}$ is always positive.
Thus, we conclude $S_3^{\prime \prime}=-1$ and the light is in the right-circularly-polarised state.
This corresponds to $S_3^{\prime}=-S_3^{\prime \prime}=1$, and such that the magnetic spin angular momentum along the principal axis seems to be preserved in spite of the large $\beta$, if the mode confinement is very strong.
The argument for the linearly polarised state is not altered by the confinement, because it is mainly affected by the choice of the frames and the sign of $\omega^{\prime}$ cannot change the direction of the polarisation, although it affects to the direction of the propagation.

\section{Conclusion}
We considered how the light will be seen in a material, if an observer is moving as fast as the phase velocity of the light.
As a specific example, we considered a graded index fibre, where the photon dispersion is massive due to the confinement of the orbital, which is quantised both for radial and angular directions.
We see that the phase velocity could change the sign, which means that the moving frame can go faster than the speed of light in a material, as evidenced by the Cherenkov radiation \cite{Cherenkov34,Frank37,Cherenkov37,Cherenkov37b,Cherenkov37c,Cherenkov86}.
We found a crossover from a red-shift to a blue-shift of light as the observer increases the speed beyond the phase velocity.
If the optical confinement in the fibre is strong, we found that anomalous Doppler effects, with the divergent phase velocity, exceeding the speed of light in a vacuum, while the group velocity is always less than $c$, confirming the causality and the validity of relativity.
We have also discussed how the polarisation state is considered in the moving frame, for which the light could be observed to be propagating to the opposite direction from the original frame.
We established that the spin operators for the light propagating in the opposite direction are described by the complex conjugate of the original spin operators, which shows the duality of the representations in $SU(2)$ Lie algebra \cite{Georgi99,Pfeifer03}.
We are not proposing to confirm this thought-experiment in reality, even if it might be possible.
Instead, we think our consideration might be useful as a platform for challenging towards the long-term mystery of what is a photon, imposed by Einstein \cite{Einstein05,Einstein05b,Lehner14}.

\section*{Acknowledgements}
This work is supported by JSPS KAKENHI Grant Number JP 18K19958.
The author would like to express sincere thanks to Prof I. Tomita for continuous discussions and encouragements.

\bibliography{Relativity}

\end{document}